\documentclass[10pt]{iopart}
\usepackage{graphicx}
\usepackage{epsfig}
\begin{document}

\jl{3}

\title[DFT study of colossal thermal expansion in Ag$_{\it3}{\textit\lbrack}$Co(CN)$_{\it6}{\textit\rbrack}$]{Origin of the colossal positive and negative thermal expansion in Ag$_3$[Co(CN)$_6$]: an {\it ab initio} Density Functional Theory study}

\author{Mark Calleja\dag, Andrew L. Goodwin\ddag, Martin T. Dove\ddag\footnote[4]{To whom correspondence should be addressed; email \tt{mtd10@cam.ac.uk}}}

\address{\dag\ Cambridge eScience Centre, Centre for Mathematical Sciences, Wilberforce Road, Cambridge, CB3 0WA, UK}

\address{\ddag\ Department of Earth Sciences, University of Cambridge, Downing Street, Cambridge, CB2~3EQ, UK}


\begin{abstract}
DFT calculations have been used to provide insights into the origin of the colossal positive and negative thermal expansion in Ag$_3$[Co(CN)$_6]$. The results confirm that the positive expansion within the trigonal basal plane and the negative expansion in the orthogonal direction are coupled due to the existence of a network defined by nearly-rigid bonds within the chains of Co--C--N--Ag--N--C--Co linkages. The origin of the colossal values of the coefficients of thermal expansion arise from an extremely shallow energy surface that allows a flexing of the structure with small energy cost. The thermal expansion can be achieved with a modest value of the overall Gr\"{u}neisen parameter. The energy surface is so shallow that we need to incorporate a small empirical dispersive interaction to give ground-state lattice parameters that match experimental values at low temperature. We compare the results with DFT calculations on two isostructural systems: H$_3$[Co(CN)$_6$], which is known to have much smaller values of the coefficients of thermal expansion, and Au$_3$[Co(CN)$_6$], which has not yet been synthesised but which is predicted by our calculations to be another candidate material for showing colossal positive and negative thermal expansion.
\end{abstract}

\pacs{61.43.Fs, 62.20.Dc, 63.50.+x}




\section{Introduction}
Recently we reported diffraction measurements showing colossal thermal expansion in the network structure of Ag$_3$[Co(CN)$_6$] \cite{Goodwinetal2008}. Within that paper we presented data showing a positive linear coefficient of expansion within the basal plane of the trigonal unit cell of $\alpha_1 = +132$ MK$^{-1}$, and a negative coefficient for the orthogonal $c$ axis of $\alpha_3 = -130$ MK$^{-1}$ (1 MK$^{-1}$ is equivalent to $1 \times 10^{-6}$ K$^{-1}$). To highlight just how large these values are, we compare values for a of number common materials in \tref{coefficients}. In the recent paper \cite{Goodwinetal2008} we briefly summarised some of results from an {\it ab initio} Density Functional Theory (DFT) study; in the present paper we report the results from this study in more detail, together with an extension to the related materials H$_3$[Co(CN)$_6$] and Au$_3$[Co(CN)$_6$].

\begin{figure}[tb]
\epsfig{file= 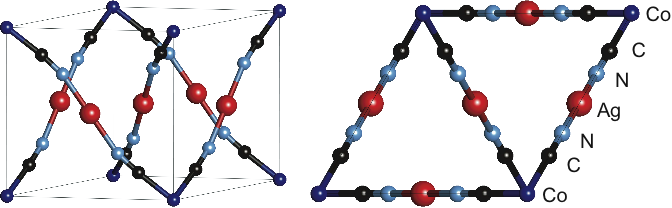,width=5in}
\caption{The crystal structure of Ag$_3$[Co(CN)$_6$], space group $P\overline{3}1m$, showing one unit cell viewed in perspective (left), and as viewed down the crystallographic $[001]$ axis (right). The atoms are labelled in the diagram. We note that the Ag atoms are the large (red) spheres at $z = 1/2$, the Co atoms are the small spheres at the origin of the unit cell (dark blue), the N atoms are the spheres closest to the Ag atoms (light blue), and the C atoms are the spheres closest to the Co atoms (black). Points to note are the existence of CoC$_6$ octahedra located on the corners of the unit cell, and the Co--C--N--Ag--N--C--Co chains lying across the set of $\{100\}$ planes. Colour is only available in the on-line version.}
\label{structure}
\end{figure}

There are two main issues involved in understanding the thermal expansion of Ag$_3$[Co(CN)$_6$]. First is to understand whether the positive expansion in the trigonal basal plane and the negative expansion in the orthogonal direction have a causal relationship. An inspection of the crystal structure, \fref{structure} suggests that the origin of the coupling between the positive and negative expansions lies in the way that the structure forms networks consisting of chains of Co--C--N--Ag--N--C--Co linkages, connecting CoC$_6$ octahedra via N--Ag--N linear groups. If the Co...Co chains remain nearly-rigid and nearly-linear, it is an inevitable consequence of the existence of this network of chains that any expansion in the basal plane has to drive a contraction of nearly equal magnitude in the orthogonal direction. However, this fact needs to be quantified. The second issue concerns what features of the material give rise to colossal values of the coefficients of thermal expansion.

\begin{table}[tb]
\caption[]{Coefficients of the linear thermal expansion tensor, $\alpha_n$, for a number of common materials. In none of these examples is the thermal expansion caused in any part by the existence of a phase transition. Units of $\alpha$ are MK$^{-1}$.}
\label{coefficients}
\begin{indented}
\item
\begin{tabular}{lccccc}
\br
Material & Temperature (K) & $\alpha_1$ & $\alpha_2$ & $\alpha_3$ & references\\
\br
Ag$_3$Co(CN)$_6$   & 10--500 & $132$ & --- & $-130$ & \cite{Goodwinetal2008} \\
\textit{d}-H$_3$Co(CN)$_6$    & 4--300 & $14.8$ & --- & $-2.4$ & \cite{RelatedSystems} \\
ZrW$_2$O$_8$           & 0.4--430 & $-9.1$ & --- & --- & \cite{Sleight_1998,Mary_1996} \\
&430--950&$-4.9$&---&---&\cite{Sleight_1998,Mary_1996} \\
low cordierite                      & 600--1050 & $6$ & $5$ & $-0.6$ & \cite{Hochella_1979}  \\
Cd(CN)$_2$                & 150--375 & $-20.4$ & --- & --- & \cite{Goodwin_2005}  \\
$\beta$-quartz                            & 900 & $1.9$ & --- & $-1.1$ & \cite{Carpenter_1998}  \\
NaCl                              & 293 & $39.6$ & --- & --- & \cite{Barron_1980} \\
\br
\end{tabular}
\end{indented}
\end{table}

Issues such as these are within the domain of atomistic simulation. In the first instance, we judged that empirical models for Ag$_3$[Co(CN)$_6$] would probably be hard to develop (for reasons that include the paucity of appropriate experimental data for tuning parameters, and the fact that the bonding is not likely to be described by the simple standard functions that would provide the necessary starting point), and thus we have started directly with DFT methods. These are described in the next section, and the main results are described in the subsequent section. We also report a set of calculations for the isostructural material H$_3$[Co(CN)$_6$], which has the same coupled positive and negative expansion but with coefficients that are closer to most other materials (\tref{coefficients}), and for Au$_3$[Co(CN)$_6$], concerning which there are currently no experimental data but which is predicted from our calculations to have similar behaviour to Ag$_3$[Co(CN)$_6$]. The final section reports a simple analysis based on the calculations, showing that the colossal thermal expansion in Ag$_3$[Co(CN)$_6$] can be accounted for with a modest value of the overall Gr\"{u}neisen parameter.

We would like to point the reader to a parallel paper that reports the results of a neutron total scattering study of Ag$_3$[Co(CN)$_6$] \cite{RMCpaper}. It contains additional discussion of the crystal structure, and presents a number of results that support the work of the present paper.

\section{Methods}
\subsection{DFT calculations}
The DFT calculations were performed using the academic version of the CASTEP code, v4.2 \cite{CASTEP} operating in energy minimisation mode. This approach uses a plane-wave representation of the electronic basis states, and represents the inner atomic electronic orbitals through the pseudopotential approximation. We used both norm-conserving (NCP) and ultrasoft (USP) pseuopotentials in various calculations, always using the generalised gradient approximation (GGA) to the exchange-correlation energy. The NCP pseudopotentials were generated using the OPIUM code \cite{opium},and the USP pseudopotentials were used as supplied with CASTEP with the exception of Co, which was kindly generated for us by Dr Victor Milman (Accelrys). In all our calculations we relaxed the crystal structure to give the lowest energy; this always involved relaxing the atomic coordinates, and in various calculations we either performed a full relaxation of the lattice parameters, or a partial relaxation, or we held the lattice parameters at fixed values.

Anticipating the results we present below, we found that DFT is unable to reproduce the correct ground-state structure of Ag$_3$[Co(CN)$_6$] for reasons that we subsequently realised point to some scientific insights. Thus we tried a number of approaches, and in this paper we report the calculations on Ag$_3$[Co(CN)$_6$] using both the NCP and USP models; the calculations on H$_3$[Co(CN)$_6$] and Au$_3$[Co(CN)$_6$] were performed using the NCP model only.  The plane-wave energy cutoffs used in the NCP and USP calculations were 680 and 470 eV respectively, obtained as good values from convergence tests. As is usual for this type of calculation, the Bloch functions were sampled over a grid of wave vectors using the standard Monkhorst--Pack grid \cite{MPgrid}, in both cases finding that converged results were obtained using a $3 \times 3 \times 3$ grid. 

The trial crystal structure used in each simulation was that determined by Pauling and Pauling \cite{Paulings} for Ag$_3$[Co(CN)$_6$], as confirmed by our recent neutron diffraction measurements \cite{Goodwinetal2008}. Where we adjusted lattice parameters (e.g.\ for running calculations across a grid of pre-set lattice parameters) we used the same set of experimental fractional atomic coordinates for the starting structure. The trial crystal structure had space group $P\overline{3}1m$ (see \tref{StructureResults}). There is no automatic method for constraining the crystal symmetry during the energy relaxation process within CASTEP, but the output results showed that the symmetry did not change through any of our relaxation calculations. The simulations on H$_3$[Co(CN)$_6$] were started with the experimental lattice parameters and fractional atomic coordinates. In the absence of experimental data, the simulations on Au$_3$[Co(CN)$_6$] used the trial structure of Ag$_3$[Co(CN)$_6$] as the starting point.

\subsection{Computational details}
The CASTEP jobs were all run on the high-throughput computing grid of the University of Cambridge, CamGrid \cite{camgridpaper,camgridurl}. This consists of a large number of 4-core (dual processors, with each processor having dual cores, and with 2 GB memory per core) shared-memory machines. Jobs were scheduled using the Condor middleware \cite{Condor}. Each job was restricted to a single node (one node being a 4-core machine), and run as a 4-processor parallel task. CASTEP was compiled with an Intel F95 compiler (v9.1), and built for the LAM flavour of MPI \cite{LAMMPI}. By selecting the \texttt{coll\_shmem} SSI module for LAM, all inter-process communications used shared memory. This avoided the computation making unnecessary use of the network stack, which would have degraded communication speed. 

\section{Results}

\begin{table}[tb]
\caption[]{Results for the calculated ground state of Ag$_3$[Co(CN)$_6$] from the different models using the CASTEP calculations: norm-conserving pseupotentials (NCP), ultrasoft pseudopotentials (USP), and compared with experimental data fo temperatures of 10 K \cite{RMCpaper} and 300 K \cite{Goodwinetal2008}. Note that the asymmetric atomic basis is defined with coordinates Co $(0,0,0)$, Ag $(1/2,0,1/2)$, C $(x,0,z)$, N $(x,o,z)$, with values for the atomic fractional coordinates $x$ and $z$ given here.}
\label{StructureResults} 
\begin{indented}
\item
\begin{tabular}{lcccc}
\br
                & Experiment (10 K) & Experiment (300 K) & NCP & USP \\
\br
$a$ (\AA)          & $6.754$    & $7.031$         & $7.764$      & $7.697$ \\
$c$ (\AA)          & $7.381$    & $7.117$         & $6.485$      & $6.681$ \\
C $x$                & $0.222$     & $0.210$         & $0.199$      & $0.198$ \\
C $z$                & $0.156$    & $0.158$         & $0.174$      & $0.167$ \\
N $x$                & $0.339$    & $0.335$         & $0.317$       & $0.317$ \\
N $z$                & $0.264$    & $0.269$         & $0.286$       & $0.278$ \\
C--N (\AA)        & $1.122$    & $1.182$          & $1.205$       & $1.179$ \\
Ag--N (\AA)      & $2.053$     & $2.012$          & $2.062$       & $2.045$ \\
Co--C  (\AA)    & $1.890$      & $1.856$          & $1.916$       & $1.889$\\
Ag--Ag  (\AA)  & $3.377$    & $3.516$         & $3.882$       & $3.849$ \\
\br
\end{tabular}
\end{indented}
\end{table}

\subsection{Ground state calculations on Ag$_3$[Co(CN)$_6$]}
The results of complete structure-relaxation calculations for both USP and NCP models are presented in \tref{StructureResults}. The striking point is that there is a substantial error in the values of the $a$ and $c$ lattice parameters -- up to 15\% -- although the Co--N, Ag--N, and C--N bond lengths are in good agreement with experiment, with errors of around 2\% (maximum error is for the Co--C distance in the NCP calculation and is 3.6\%). The discrepancies in individual bond lengths are typical for this type of calculation (the community experience is that GGA usually leads to a small over-estimate of bond lengths), but the large errors in calculated lattice parameters are not expected for the methods we used. We believe that the large discrepancy between experiment and calculations is actually providing some critical information about the physics and chemistry of Ag$_3$[Co(CN)$_6$], suggesting that an important feature is missing from the DFT calculations. Anticipating results described later in this paper, we note that our calculations on H$_3$[Co(CN)$_6$] gave results that are much closer to experiments (within the usual GGA level of agreement), suggesting that the discrepancies are due to some specific feature of the Ag cations.

Given that the DFT calculations are giving reasonable results for the individual bond lengths, we repeated the calculations in which the lattice parameters are held at their experimental values and atomic coordinates were allowed to relax.  For the NCP and USP models,  the energy differences are $0.111$ and $0.169$ eV per formula unit respectively ($0.037$ and $0.056$ eV per Ag atom respectively). This is actually quite a small difference, and suggests that whatever is missing from the DFT calculations is not large, but nevertheless appears to be significant in the present case because overall energy differences are small. It is also interesting to note that the DFT results do not show any appreciable rearrangement of the atomic charges or change in the bonding characteristics when changing the lattice parameters --- the calculated Mulliken charge distribution and Mulliken bond orders obtained directly from the CASTEP calculations \cite{Mcharge1,Mcharge2} are given in \tref{Charges}, and show no significant changes with large changes in lattice parameters. 

\begin{table}[tb]
\caption[]{Mulliken atomic charges ($Q$) and Mulliken bond orders (BO) calculated for Ag$_3$[Co(CN)$_6$] (first four columns of data) and H$_3$[Co(CN)$_6$] and Au$_3$[Co(CN)$_6$] (last two columns). NCP and USP correspond to the use of norm-conserving pseupotentials and ultrasoft pseudopotentials respectively, and MIN and EXP refer to the lattice parameters of Ag$_3$[Co(CN)$_6$] after minimisation and with the experimental values respectively.}
\label{Charges} 
\begin{indented}
\item
\begin{tabular}{lcccccc}
\br
 & NCP\_MIN & USP\_MIN & NCP\_EXP & USP\_EXP & H$_3$[Co(CN)$_6$] & Au$_3$[Co(CN)$_6$] \\
\br
$Q_{\mbox{\tiny{Co}}}$        & $1.13$     & $1.15$       & $1.16$      & $1.19$     & $1.32$  &  $1.18$ \\
$Q_{\mbox{\tiny{Ag/H/Au}}}$    & $0.68$     & $0.61$       & $0.65$      & $0.58$     & $0.36$ & $0.54$ \\
$Q_{\mbox{\tiny{C}}}$          & $-0.01$    & $-0.02$      & $-0.01$     & $-0.02$     & $0.05$ &  $0.00$ \\
$Q_{\mbox{\tiny{N}}}$          & $-0.52$    & $-0.48$      & $-0.51$     & $-0.47$     & $-0.45$ &  $-0.47$ \\
BO Co--C                               & $0.25$        & $0.20$      & $0.25$     & $0.19$     & $0.22$  &  $0.25$  \\
BO C--N                                 & $1.80$        & $1.84$      & $1.80$     & $1.84$     & $1.77$ &  $1.81$ \\
BO Ag/H/Au--N                     & $0.33$        & $0.35$      & $0.34$     & $0.37$     & $0.43$  &  $0.41$ \\
BO Ag--Ag/H--H/Au--Au       & $-0.01$        & $-0.03$          & $0.00$       & $-0.02$     & $0.00$ &  $0.00$  \\
\br
\end{tabular}
\end{indented}
\end{table}

\begin{figure}[tb]
\epsfig{file= 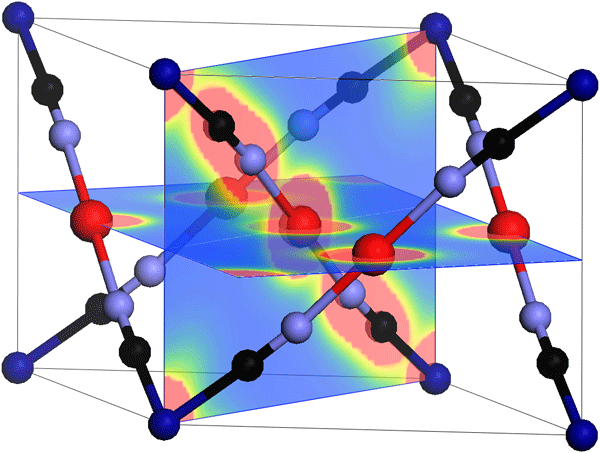,width=4in}
\caption{Electron density contour plots for two planes in the crystal structure of Ag$_3$[Co(CN)$_6]$. The vertical plane $(1\overline{1}0)$ contains the Co--C--N--Ag linkages and shows the strong covalent bonding within the C--N group and additional covalent Co--C and Ag--N bonds as electron density between the bonded atoms. The horizontal plane ($(001)$ plane at $z=1/2$) contains the nearest-neighbour Ag cations, and shows no electron density between these atoms. The labelling of the atoms is as given in \fref{structure}. Colour is only available in the on-line version.}
\label{edensity}
\end{figure}

Our hypothesis is that the large discrepancies between the experiment and calculated ground state structures can be traced to the Ag...Ag nearest-neighbour (non-bonded) distance, which is half the value of the $a$ lattice parameter. This is longer in the calculation than experiment by $0.5$ \AA. The Ag...Ag distance at low temperature -- $3.38$ \AA\ -- is of the same size as twice the Van der Waal's radius ($1.72$ \AA), and thus might be considered to be rather short in view of the fact that this is a charged cation (DFT charge value is of order $0.6$ positive electron units). The electron density obtained from the NCP calculations is presented in \fref{edensity}. This shows that there is no covalent bonding between the neighbouring Ag cations, although covalent bonding is seen along the Co--C--N--Ag chains. Thus we believe that the the expansion in the basal plane in the calculation relative to the experimental structure can be understood as arising from a repulsion between the neighbouring Ag atoms, or equivalently, due to the lack of an additional attractive interaction in the model. In short, there is some aspect of the Ag...Ag interactions that is not properly reflected in the DFT calculations, albeit one that will not give large changes in energy (recall the comparison between the calculations with complete structure relaxation and those held at the experimental lattice parameters above).

At this point we note the work of Pyykk\"{o} and co-workers on metallophilic interactions involving group 11 elements \cite{Pyykko97, Pyykko02, Pyykko04}. These appear to arise due to electron correlation effects that are not included within the DFT approach \cite{Du_Smith}, and can be represented as an enhanced dispersive interaction. Based on this, we have attempted to correct for the discrepancies between calculation and experiment in Ag$_3$[Co(CN)$_6]$ by adding a \textit{post hoc} dispersive interaction between the Ag atoms of the form $-Cr^{-6}$. We find that the minimum energy state with lattice parameters closest to the experimental low-temperature values can be achieved using a value of $C$ of $80$ eV \AA$^6$ per atom pair. We should remark that in general \textit{post hoc} corrections would not be expected to give consistent results, because we minimise the atomic coordinates without the correction. However, in the present case the calculations show that the bond lengths remain more-or-less constant independent of the values of the $a$ and $c$ lattice parameters, and thus we do not expect the inclusion of the \textit{post hoc} correction to significantly affect bond lengths. Of course, we could have also included dispersive interactions involving the C and N atoms, but based on our calculations of H$_3$[Co(CN)$_6$] presented below we have compelling evidence that the Ag...Ag interactions are the critical ones here. In any case we anticipate that any effects of the C and N atoms will be incorporated into the Ag...Ag interaction we included through the value of $C$ we used. Thus we conclude that Ag...Ag interactions that are not captured within the DFT method are required to obtain the correct ground state of Ag$_3$[Co(CN)$_6$]. However, these are not large and are only noticeable because of the high degree of flexibility of the network that defines the crystal structure. We show below that this additional interaction actually softens the energy surface, and is implicated in the ability of this material to show colossal positive and negative thermal expansion.

\begin{figure}[tb]
\epsfig{file=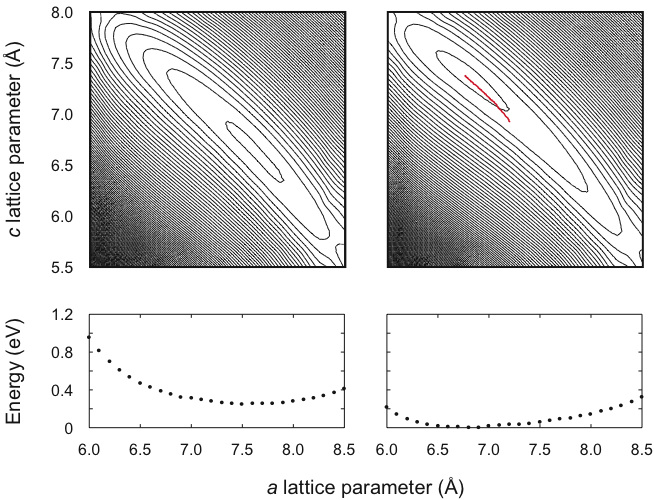,width=5in}
\caption{Contour plots for the grid of calculations for Ag$_3$[Co(CN)$_6$] holding $a$ and $c$ at fixed values, obtained using with norm-conserving pseudopotentials without (\textit{top left}) and with (\textit{top right}) the {\it post hoc} dispersive interaction. The line of points (red in on-line version) show the locus of experimental data.  The lower curves trace the energy of the locus of points along the bottom of the valley in the contour plots.}
\label{Contours}
\end{figure}

\subsection{Energy surface for Ag$_3$[Co(CN)$_6$]}
In \fref{Contours} we plot the energy surface computed for a grid of fixed values of $a$ and $c$ for the model with norm-conserving pseudopotentials, allowing relaxation of the atomic coordinates for each calculation. The results are shown with and without the {\it post hoc} dispersive interaction. The key result is that the energy surface consists of a very steep valley with an extremely shallow `floor'. The \textit{post hoc} additional dispersive interaction has little effect on the overall shape of the valley, but shifts the minimum point along the floor. \Fref{Contours} also contains the locus of experimental data for $a$ and $c$ for various temperatures \cite{Goodwinetal2008}; the position of this locus is very close to the calculated valley floor, confirming that the DFT calculations are capturing the correlation between the values of $a$ and $c$ seen in experiment. In \fref{Contours} we also plot the energy from from the locus of minima points obtained by holding fixed values of $a$ and $b$ and allowing $c$ to relax.

\begin{figure}[tb]
\epsfig{file=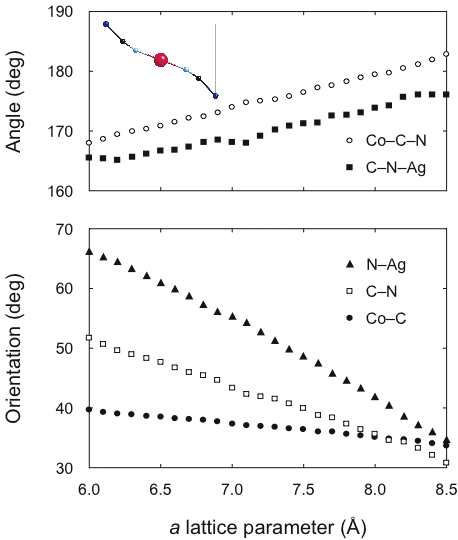,width=4in}
\caption{Plots of the angles (\textit{top}) and orientations with respect to the $(001)$ plane (\textit{bottom}) of the nearest-neighbour bonds in Ag$_3$[Co(CN)$_6$], calculated using fixed values of $a$ and allowing $c$ and the atomic coordinates to relax. The angles are defined such that the angles in the cartoon, which corresponds to the structure for $a = 6.0$ \AA, are both acute, and the orientations are given with respect to the horizontal line (see cartoon).}
\label{BondAngles}
\end{figure}

\subsection{Structure analysis}
We note here that the Co--N, C--N and N--Ag bond lengths do not change significantly across the range of the calculations. Maximum changes in bond lengths for the Co--N and N--Ag bonds are 1\% and 0.8\% respectively for the range of calculations in which we held the $a$ lattice parameter fixed at values between 6 and 8.5 \AA, and allowed relaxation of the $c$ lattice parameter and of the atomic coordinates. The change in the C--N bond length across the range of calculations is even smaller, 0.1\%.

Thus the largest changes in the structure for large changes in the lattice parameters are in the orientations of the bonds, or equivalently in the bond angles. In \fref{BondAngles} we plot the calculated Co--C--N and C--N--Ag bond angles (\textit{top}), and the calculated orientations of the Co--C, C--N and N--Ag bonds with respect to the $(001)$ plane (\textit{bottom}), all as a function of the lattice parameter $a$ obtained with relaxation of $c$ and the atomic coordinates. Note that a linear Co--C--N--Ag chain will correspond to the three orientations being equal. Thus we note that for large values of the $a$ lattice parameter we have a greater tendency towards forming linear chains, but on decreasing the value of $a$ we see significant crumpling of these chains. The orientation of the Co--C bond is actually determined by the shape of the CoC$_6$ octahedron, and the fact that this changes the least -- by no more than $5^\circ$ -- is consistent with the fact that the CoC$_6$ octahedra retain their basic shape across the range of calculations. The C--Co--C angle (not shown in \fref{BondAngles}) varies in a more-or-less linear manner from $84^\circ$ at $a = 6$ \AA\ to $92^\circ$ at $a = 8.5$ \AA. In order to accommodate the large decrease in the value of $a$, and the concomitant increase in the value of $c$, it is essential that the chain be aligned closer towards the vertical direction. Thus there is a clear need for the chain to buckle, as seen in \fref{BondAngles}. It is interesting to note that the two chain bond angles change by similar amounts (\fref{BondAngles} \textit{top}), so that the N--Ag bond changes its orientation by the greatest amount (\fref{BondAngles} \textit{bottom}), rather than the buckling being accommodated more by bending of one bond angle than the other (note that the N--Ag--N linkage is constrained to be linear).

\begin{figure}[tb]
\epsfig{file=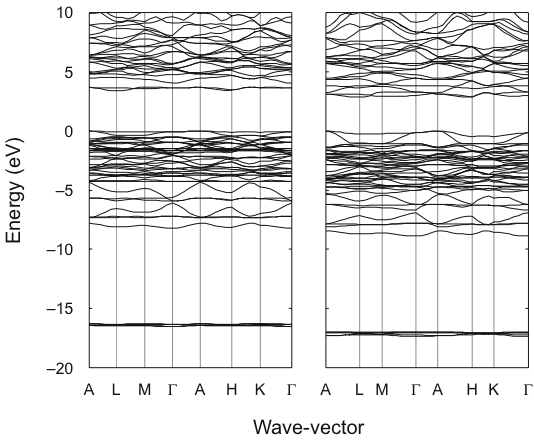,width=5in}
\caption{Plots of the DFT band structure of Ag$_3$[Co(CN)$_6$]  computed at the values of lattice parameter $a = 7.7652$ \AA\ of the minimum DFT energy structure (\textit{left}) and for $a = 6.0$ \AA\ (\textit{right}; lattice parameter $c$ and atomic coodinates were at relaxed values).}
\label{bandstructure}
\end{figure}

\subsection{DFT Band structure}
The DFT band structure of Ag$_3$[Co(CN)$_6$] is showin \fref{bandstructure} for two different values of lattice parameter $a$, namely 6.0 \AA\ and 7.7652 \AA\ (the lattice parameter of the DFT minimum energy structure), computed on structures for which the lattice parameter $c$ and the atomic coordinates had been relaxed. Two points should be noted from the data. First, the DFT band gap -- the existence of which is consistent with the experimental observation that  Ag$_3$[Co(CN)$_6$] is an insulator -- increased from 2.84 eV to 3.37 eV on increasing $a$ from 6.0 \AA\ to the value of the DFT energy minimum structure. The band gap increases further to 3.70 eV on increasing $a$ to 8.4 \AA. Although DFT is known to significantly underestimate the true band gap, trends in the behaviour of the band gap may be considered to be qualitatively reliable. Thus, whilst the changes in the DFT band gap are not large, there is nevertheless an appreciable lowering of the band gap as the Ag atoms come closer together, and as the C...C chains become more vertically inclined. The second point to note is that the splitting of the top two valence bands increases significantly on decreasing $a$, with one band developing a greater variation with wave vector. These two bands are degenerate at the $\Gamma$ and A ($\mathbf{k} = 0,0,1/2$) points (and along the $\Gamma$--A line), and there is maximum splitting along the L--M and H--K lines (L, M and H are $\mathbf{k} = 1/2,0,0$, $1/2,0,0$ and $1/3,1/3,1/2$ respectively \cite{Bilbao}).

\subsection{Aside: orientation of the CN molecular ion}
From a purely electrostatic viewpoint, the Co cations might be expected to have N rather than C as its neighbouring atom, particularly as the N component of the CN$^-$ molecular ion contains most of the charge (as seen in \tref{Charges}). Indeed, all known structural analogues, such as La[Ag(CN)$_2$]$_3$ and KCo[Ag(CN)$_2$]$_3$, have the CN molecular ion rotated by $180^{\circ}$ relative to Ag$_3$[Co(CN)$_6$]. This raises the question of whether the orientation of the CN molecular ion in Ag$_3$[Co(CN)$_6$] is in a genuine equlibrium arrangement, or whether this structure is the metastable result of the growth process. We therefore calculated the energy for the case where we swapped the positions of the C and N atoms, giving Co--N--C--Ag--C--N--Co chains instead of the Co--C--N--Ag--N--C--Co chains found in the experimental structure. Our calculations showed that the the experimental structure has the lower energy difference, with an energy difference of $1.03$ eV per unit cell from the DFT calculations with the norm-conserving pseudopotentials.

\subsection{Calculations on H$_3$[Co(CN)$_6$]}
To provide a comparison, we have performed calculations (using the NCP model) on the isostructural material H$_3$[Co(CN)$_6$]. This is a useful comparison because experimental data for H$_3$[Co(CN)$_6$] show that that the thermal expansion is much smaller in this material (see \tref{coefficients}). We used the same structure as for Ag$_3$[Co(CN)$_6$], with the H lying exactly half way between the neighbouring N atoms. There has been some discussion in the literature concerning whether the positions of the H atoms might be disordered. In particular, the structure reported by 
Haser \textit{et al} \cite{Haseretal} has two sites for each H atom. However, our own recent neutron powder diffraction studies \cite{RelatedSystems} have found no evidence for positional site disorder, with the best structure refinements having the H atoms in ordered positions, albeit with evidence for transverse vibrations as might be expected.

\begin{table}[tb]
\caption[]{Results for the calculated ground state of H$_3$[Co(CN)$_6$] and Au$_3$[Co(CN)$_6$]; atomic coordinates follow the definition in \tref{StructureResults} with H and Au replacing Ag, and experimental results for deuterated-H$_3$[Co(CN)$_6$] were obtained at a temperature of 8 K \cite{RelatedSystems}.}
\label{StructureResults2} 
\begin{indented}
\item
\begin{tabular}{lcccc}
\br
 & \textit{d}-H$_3$[Co(CN)$_6$] experiment & H$_3$[Co(CN)$_6$] calculation & Au$_3$[Co(CN)$_6$] calculation\\
\br
$a$ (\AA) & $6.412$ & $6.545$ & $7.990$ \\
$c$ (\AA) & $5.718$ & $5.678$ & $6.193$ \\
C $x$ & $0.238$ & $0.238$ & $0.196$ \\
C $z$ & $0.194$ & $0.198$ & $0.180$ \\
N $x$ & $0.376$ & $0.377$ & $0.312$ \\
N $z$ & $0.322$ & $0.324$ & $0.292$ \\
C--N & $1.151$ & $1.161$ & $1.163$ \\
H/Au--N & $1.291$ & $1.283$ & $1.978$ \\
Co--C & $1.885$ & $1.917$ & $1.918$ \\
H--H/Au--Au & $3.206$ & $3.272$ & $3.995$ \\
\br
\end{tabular}
\end{indented}
\end{table}

The results of the DFT calculations are summarised in \tref{StructureResults2}. The agreement between the structures calculated by DFT and obtained from experiment is reasonable, and in line with the accuracy that is typical for DFT calculations: the errors on the values of $a$ and $c$ are $+2$\% and $-0.6$\%. In light of the above discussion on the role of Ag...Ag interactions, it is interesting to note this good agreement, which indirectly confirms the important role of Ag...Ag interactions in determining the ground state of Ag$_3$[Co(CN)$_6$].

\begin{figure}[tb]
\includegraphics[width=5in,clip=true]{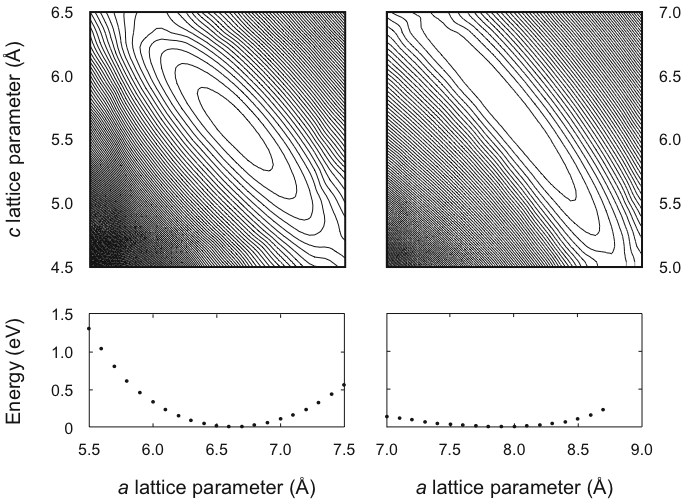}
\caption{Contour plot for the DFT calculations for H$_3$[Co(CN)$_6$] (\textit{top left}) and Au$_3$[Co(CN)$_6$] (\textit{top right}) with the trace of the energy of the locus of points along the bottom of the valley in the contour plot (\textit{bottom}). It is interesting to compare these plots with the data presented in \fref{Contours}.}
\label{Contours2}
\end{figure}

In \fref{Contours2} (left side) we show the contour plot for H$_3$[Co(CN)$_6$] analogous to the plot shown for Ag$_3$[Co(CN)$_6$] in \fref{Contours}, together with the locus of the minima points along the valley in the energy surface. In comparison with Ag$_3$[Co(CN)$_6$] we see a similar steep valley in the energy surface that provides the correlation between inverse changes in lattice parameters $a$ and $c$. However, the changes in energy along the floor of the valley are significantly larger than in Ag$_3$[Co(CN)$_6$]. Given that this is the most significant difference between the calculations for the two materials, we surmise that it is closely associated with the differences in values of the coefficients of thermal expansion.

\subsection{Calculations on Au$_3$[Co(CN)$_6$] }
Finally, we also report calculations of the ground-state structure of Au$_3$[Co(CN)$_6$], \tref{StructureResults2}, and a calculation of the energy surface is shown in \fref{Contours2}. We believe that Au$_3$[Co(CN)$_6$] has yet to be synthesised, so these results are predictive. The energy surface is very similar to that of Ag$_3$[Co(CN)$_6$] (\fref{Contours}), and thus we can predict that the thermal expansion will also show a similar correlation between $a$ and $c$. Moreover, the variation of energy along the floor of the valley in the energy surface is very close in size to that of Ag$_3$[Co(CN)$_6$] -- and is much smaller than that of H$_3$[Co(CN)$_6$] -- so we anticipate that Au$_3$[Co(CN)$_6$] will large values of its coefficients of thermal expansion similar in size to those of Ag$_3$[Co(CN)$_6$]. Because there are no experimental data on this structure, it is impossible to judge the extent to which an additional dispersive Au...Au interaction is required to give a complete description of the energy surface, but the work of Pyykk\"{o} and co-workers \cite{Pyykko97,Pyykko02,Pyykko04} suggest that there should be significant metallophilic interactions as in Ag$_3$[Co(CN)$_6$]. Furthermore, the calculated value of the $a$ lattice parameter is much larger (of order $1.43$ \AA) than the experimental value for the isostructural La[Au(CN)$_2$]$_3$, where $a = 6.664$ \AA\ at 213 K \cite{Dalton}. This suggests that it is highly likely that the behaviour of Au$_3$[Co(CN)$_6$] and analogues such as La[Au(CN)$_2$]$_3$ will show colossal positive and negative thermal expansion.

\section{Analysis}
The DFT results shown in \fref{Contours} point clearly to the existence of an energy surface in the space defined by the $a$ and $c$ lattice parameters that consists of a valley with very steep sides but only a shallow slope along the locus of minima points. The shape of the energy surface thus explains the origin of the negative thermal expansion along $c$ for a positive expansion along $a$, consistent with the intuition from the structure outlined in the Introduction.

We now consider the question of why the coefficients of thermal expansion of Ag$_3$[Co(CN)$_6$] have colossal magnitudes. In our analysis we anticipate that the thermal expansion can be driven as a normal anharmonic phonon process with a very shallow energy surface. We write the free energy in standard form \cite{SD} as 

\begin{equation}
F = 3 N k_{\mathrm{B}}T \ln \left( \frac{\hbar \omega}{k_\mathrm{B} T} \right) + \frac{1}{2}\frac{\partial ^2 E}{\partial V^2}(V - V_0)^2
\end{equation}

\noindent where $V$ is the volume occupied by $N$ atoms, $V_0$ is the corresponding volume at the minimum-energy configuration, $E$ is the potential energy of $N$ atoms, and the phonon free energy has been expressed in the high-temperature limit ($k_{\mathrm{B}}T > \hbar \omega$) where $\omega$ represents an average phonon angular frequency. Minimisation with respect to volume yields

\begin{equation}
V = V_0 + 3N  \gamma B k_{\mathrm{B}}T = V_0(1 + \beta T)
\end{equation}

\noindent where $\gamma$ is the normal overall Gr\"{u}neisen parameter, and the bulk modulus $B = (V_0 \partial^2 E / \partial V^2)^{-1}$. The equation defines $\beta = 3NBk_{\mathrm{B}}\gamma/V_0$. Our approach here is to use the energy calculations and experimental data for $\beta$ (obtained as the trace of the linear thermal expansion coefficients) to obtain a value for $B\gamma$, and from an estimate for $B$ (this can only be an estimate because of the use of the \textit{post hoc} correction for the dispersive energy) to obtain an estimate of $\gamma$. Thus we obtain $B\gamma = 47.3 \times 10^{-12}$ Pa$^{-1}$ from the experimental data. The origin of the colossal values of the coefficients of thermal expansion can be expressed in terms of whether Ag$_3$[Co(CN)$_6$] has an unusually large value of $B$ or $\gamma$ or both.

From the energy calculations we obtained a fitted value for $B^{-1} = 9.9$ GPa, which would yield a value of $\gamma = 0.47$. We stress that our fitted value of $B^{-1}$ is hard to estimate accurately from our DFT data because, being a small value, it is affected by very small errors in the calculations. However, it is nevertheless clear that the value for $B^{-1}$ is small (by comparison, values of $B^{-1}$ for NaCl, CaF$_2$ and cordierite are 24.4, 82.7 and 129 GPa respectively), so that $B$ is large, and this had been anticipated from the significant strain broadening seen in diffraction experiments \cite{Goodwinetal2008}. In fact, the value of $B$ is so large that we have obtained a fairly typical estimate (in fact, on the low side of normal) for the value of $\gamma$ rather than a particularly large value.

We note that we do not anticipate a large value for $\gamma$ for two reasons: first, because we do not anticipate the bond-stretching vibrational frequencies to change significantly with volume given that the bond lengths and bond orders do not change in the DFT calculation; and second, because negative contributions to the overall Gr\"{u}neisen parameter will also decrease its value. These will arise from dynamic bond-bending motions along the Co--C--N--Ag--N--C--Co chains, which will have the effect of pulling the end Co atoms closer together.\footnote{An analysis of the numbers of structural constraints and degrees of freedom shows that there 15 more degrees of freedom than constraints per formula unit. This imbalance gives rise to rotational motions of stiff CoC$_6$, CN and AgN$_2$ units that will cause a shortening of the Co...Co chain. These motions correspond to the Rigid Unit Modes that are known to give negative contributions to the overall Gr¬{u}neisen parameter \cite{Goodwin_2005,Patrick,Hammonds,Pryde}.} We note that these motions have been observed in our RMC study. The final value of $\gamma$ will be a balance between motions that expand the volume and those that want to pull the structure inwards.

What this analysis shows is that the colossal values of the coefficients of thermal expansion can be explained reasonably well with a normal, albeit slightly small, value of the phonon Gr\"{u}neisen parameter, and that the values of the coefficients of thermal expansion are so large because of the incredibly shallow energy surface along the bottom of the valley.

We repeat this analysis for our results for H$_3$[Co(CN)$_6$]. We obtain $B\gamma = 8.36 \times 10^{-12}$ Pa$^{-1}$ from experimental data. Our calculated energy surface gives $B^{-1} = 56.4$ GPa, from which we obtain $\gamma = 0.47$. This value of $\gamma$ is very similar to that of Ag$_3$[Co(CN)$_6$]. The principle difference between the thermal expansion of the two materials is thus due entirely to the different values of $B$, which is seen in the differences between the energy surfaces. 

Finally, from the shapes of the energy surfaces, and from the result that both Ag$_3$[Co(CN)$_6$] and H$_3$[Co(CN)$_6$] have normal values for $\gamma$, we make the prediction that Au$_3$[Co(CN)$_6$] will also show the same behaviour for its thermal expansion as Ag$_3$[Co(CN)$_6$], as also may other isostructural materials containing silver or gold. The main challenge involved in studying this compound experimentally is the limited aqueous chemistry of the Au$^+$ cation, which prevents a simple modification of the synthetic route for Ag$_3$[Co(CN)$_6$].

\section{Conclusion}
In this paper we have used DFT calculations to demonstrate that the coupled positive and negative thermal expansion in Ag$_3$[Co(CN)$_6$] and H$_3$[Co(CN)$_6$] arises from the existence of steep valley in the appropriate energy surface that can be traced to the existence of chains of atoms within the crystal structure that can flex like a garden trellis. The floor of the valley is particularly shallow in the case of Ag$_3$[Co(CN)$_6$], which gives rise to a large value of the volume compressibility (small value of the bulk modulus). This factor alone gives rise to the existence of colossal values of the coefficients of thermal expansion, with the values of the Gr\"{u}neisen parameters appearing to be normal. This work has pointed to the important role of Ag...Ag metallophilic interactions, although we cannot claim to have arrived at a complete understanding of how these give rise to such a shallow floor of the energy surface. More work is needed in this regard. Moreover, we have predicted that Au$_3$[Co(CN)$_6$] will show similar behaviour to Ag$_3$[Co(CN)$_6$], and we anticipate that other materials with metallophilic interactions will also display extreme behaviour such as collosal thermal expansion.

\ack 
The authors would like to acknowledge useful discussions with Dr Dan Wilson (Frankfurt) and Prof David Keen (ISIS), and help from Dr Helen Chappell (Cambridge) in the use of visualisation tools. We thank Dr Victor Milman (Accelrys) for supplying the Co USP used in this work. The calculations were performed using the computational resources of CamGrid \cite{camgridpaper,camgridurl}. 


\end{document}